\documentclass[aps,prl,twocolumn,groupedaddress,showpacs]{revtex4-1}

\usepackage{graphicx}
\begin{document}

%\preprint{}

\title{Onset of Plasticity in thin Polystyrene Films}

\author{Bekele J. Gurmessa}
\author{Andrew B. Croll}
%\thanks{}
\affiliation{Department of Physics, North Dakota State University, Fargo, ND, USA}
\email{andrew.croll@ndsu.edu}


\date{\today}

\begin{abstract}
Polymer glasses have numerous advantageous mechanical properties in comparison to other materials.  One of the most useful is the high degree of toughness that can be achieved due to significant yield occurring in the material.  Remarkably, the onset of plasticity in polymeric materials is very poorly quantified, despite its importance as the ultimate limit of purely elastic behavior.  Here we report the results of a novel experiment which is extremely sensitive to the onset of yield and discuss its impact on measurement and elastic theory.  In particular, we use an elastic instability to locally bend and impart a \textit{local} tensile stress in a thin, glassy polystyrene film, and directly measure the resulting residual stress caused by the bending.  We show that plastic failure is initiated at extremely low strains, of order $10^{-3}$ for polystyrene.  Not only is this critical strain found to be small in comparison to bulk measurement, we show that it is influenced by thin film confinement - leading to an increase in the critical strain for plastic failure as film thickness approaches zero.
\end{abstract}

\pacs{ 46.25.-y, 81.05.Lg, 62.20.Fe, 82.35.Lr  }	% PACS, the Physics and Astronomy

\maketitle

%\section{}
%\subsection{}
%\subsubsection{}

Thin polymer films are materials that are uniquely suited for countless goals, finding use in studies ranging from fundamental polymer physics~\cite{Brown1996,Si2005}, to  elasticity~\cite{Cerda2003,Stafford2004, Witten2007, Huang2007,  Pocivavsek2008,  Davidovitch2011}, and glass-phase physics~\cite{Keddie1994, Ellison2003, Fakhraai2008, OConnell2005, Pye2011,Bodiguel2006,Green2002}.   Polystyrene is often used as a model polymeric material because thin films are easily produced and manipulated, they are relatively rigid, and polystyrene is a comprehensively studied glass-forming polymer with well known \textit {bulk} material properties.  However, as the thickness of a polymer film is reduced to nanoscopic dimensions it is now well established that the glass transition is altered~\cite{Keddie1994, Ellison2003, Fakhraai2008, OConnell2005, Pye2011,Bodiguel2006,Green2002}, the number of entanglements between chains are reduced~\cite{Brown1996,Si2005}, and mechanical properties can change~\cite{Lee2012,Bohme2002}.  Remarkably few studies have considered how failure processes might be altered in thin films, or how plasticity might alter a films response to other stimuli.

Many researchers have recently focused on a surface instability known as wrinkling that occurs when a thin polymer film (modulus $\bar{E}_{f}$, dimensions $L \times L$ and thickness $h$) is bound to a stress free elastic substrate (modulus $\bar{E}_{f}$) and is subjected to a compressive uniaxial stress \cite{Bowden1998}.  When stress is applied, the film is initially compressed in plane but quickly reaches a critical point above which the film buckles out of plane~\cite{Huang2007, Stafford2004}.  In the limit of small strain, $\epsilon$,  a balance between the energy associated with the bent plate $U_{B}\sim \epsilon L^{2} \bar{E}_{f} h^3 / \lambda^2$ and that of the stretched substrate $U_{S}\sim \epsilon L^{2} \bar{E}_{S} \lambda$ reveals the utility of the system - the pattern is directly related to the thin film material properties (minimization gives $\lambda = 2\pi h (\bar{E}_{f}/3\bar{E}_{s})^{2/3}$)~\cite{Cerda2003,Groenewold2001,Bowden1998,Stafford2004}.  This simple analysis was developed for the ill defined 'near threshold' regime, although for reasons that are not yet clear the analysis seems to be accurate in many situations that are clearly no longer small perturbations~\cite{Huang2007}.  Larger displacements (the so called far from threshold regime) currently challenges elastic theory for a solution, which has prompted significant experimental and theoretical efforts~\cite{Pocivavsek2008, Davidovitch2011, Holmes2010, Ebata2012}.  The far from threshold work often investigates aspects of stress focusing~\cite{Witten2007}, deformations in which the stress distributions become strongly heterogeneous and notably similar to more general types of material failure (e.g. fracture).  

Reaching the far from threshold regime typically requires the application of large stresses.  High stress also leads to failure, for example the nucleation of fractures (or shear deformation zones and crazes) in a wrinkled film~\cite{Lee2012} or the formation of delaminations if the wrinkling energy $U_{w} \sim \epsilon L h \bar{E}_{f}^{1/6} \bar{E}_{s}^{5/6}$ exceeds the adhesion energy between the film and the substrate ($U_{a}\sim G_{c} w \ell$, where $w \ell$ is the area of the delamination and $G_{c}$ is the energy release rate)~\cite{Ebata2012, Mei2011, Audoly2010}.  More importantly, high stress may push the film beyond the material limits of linear elasticity, where plasticity will greatly affect certain aspects of stress focusing~\cite{Tallinen2009}.   Therefore, for both practical and fundamental reasons, it is critical to understand where and how plastic deformation occurs in thin polymer films.  In this letter we describe a sensitive measurement of the onset of plasticity in thin polystyrene films, an important material property which has thus far been experimentally overlooked.  Not only do we show that the critical strain for onset of plastic deformation, $\epsilon_{p}$, in a thin polymer film is considerably lower than what is expected from bulk measurements, but we show that the critical point is affected by thin film confinement in a manner that is not related to entanglement density.  

\begin{figure}
\centering
\includegraphics[width=0.45\textwidth]{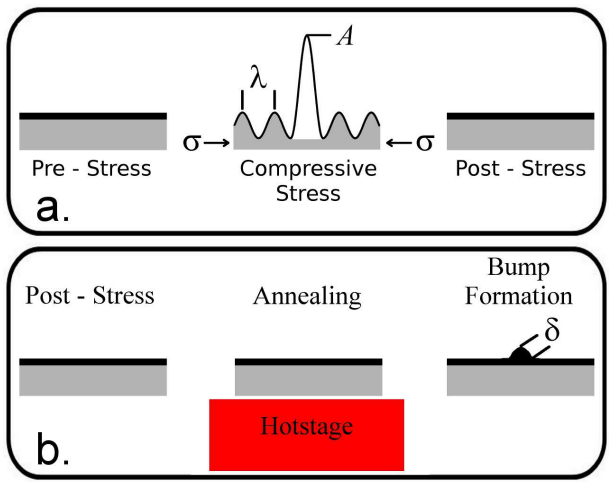}
%\resizebox{0.9\columnwidth}{!}{schem8.eps}
\caption{Schematic illustrating the two stage experiment.  (a) Stage 1: a thin polymer film is placed on a soft elastomer substrate, uniaxial compression is added and finally removed.  The sample is left in a measurably flat state.  (b) Stage 2: the sample, after all applied stress has been removed, is placed on a  hot-stage and annealed.  During annealing, ``bumps'' form at locations correlated with high curvature during the compression stage of the experiment.}\label{scheme}
\end{figure}

Our experiment is intentionally simple: we apply and then remove a compressive stress from thin, glassy polystyrene (PS) films bound to an elastomer substrate - in essence a double step strain experiment (see Fig.~\ref{scheme}).  Under compression, the film buckles forming a wrinkle pattern that evolves to a complex collection of delamination blisters and wrinkles as compression is increased~\cite{Mei2011, Ebata2012}.  Upon release of stress, the topography of regions that were highly curved are measured and found to be flat on the order of the surface roughness ($\sim 1$~nm).  There are two possible outcomes from such an experiment: 1) if the deformation is purely elastic the film is returned to its initial (unstressed) configuration or 2) if there is a plastic component to the films response, the film will be left in a higher stress configuration (we remain far below the films glass transition temperature, $T_{g}$, during application of step strain).  The experimental difficulty lies in determining whether the film is in state 1 or state 2 after the experiment.  Here we make this distinction by subsequently annealing the film above its glass transition temperature ($T_{g} \sim 370$~K for bulk PS) - if the film is in state 1 it will not appreciably change, however if the film has stored stress (state 2) the now fluidized film must flow in response.  If the plastic stress is uniform and isotropic the response will only appear in the dynamics of the flow~\cite{Reiter1994, Reiter2005, Chung2009}, or as a small change in film thickness.  The onset of plastic flow would be difficult to measure in both these schemes.  Hence, we add a further refinement to the experiment and use the alternating positive and negative curvature of the wrinkle topography to create a \textit{local} bending stress in the sample.

\begin{figure}
\centering
\includegraphics[width=0.45\textwidth]{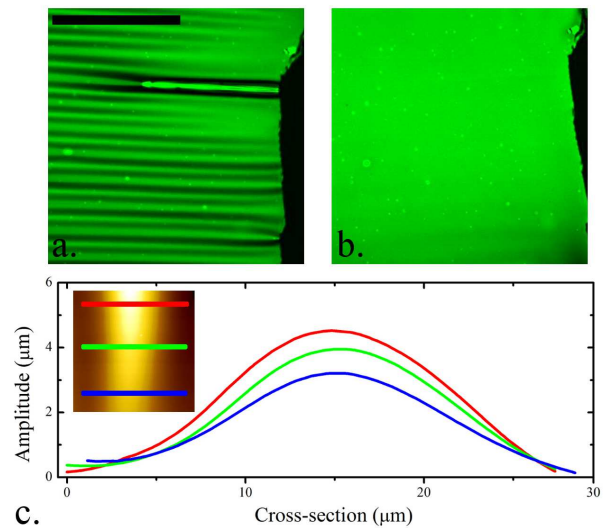}
\caption{ Typical data from stage 1 of the experiment. a. A confocal microscope image of wrinkles (low curvature) and a delamination (high curvature) near a sample edge. Scale bar indicates $32$~$\mu$m.  b. The same location after removal of stress - there is no observable change in the film.  c. Cross sections through the delamination illustrating its variable amplitude (inset: an AFM micrograph of a segment of delamination).}\label{before}
\end{figure}

\begin{figure}
\centering
\includegraphics[width=0.45\textwidth]{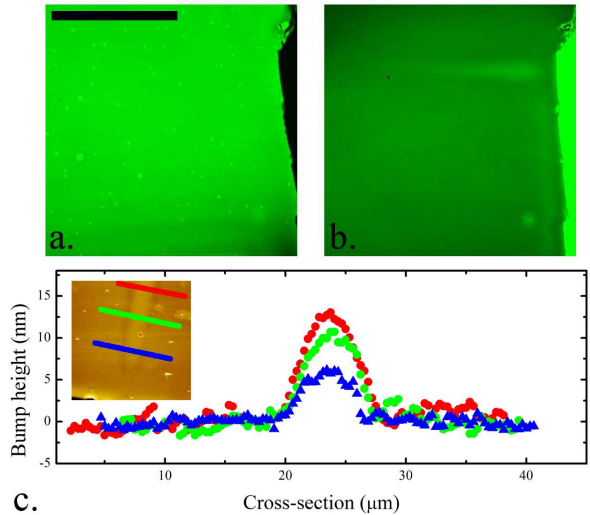}
\caption{ Typical data from the second stage of the experiment. (a) the same film shown in Fig.~\ref{before} after compression.  Scale bar indicates $32$~$\mu$m.  (b) After annealing at 140~$^\circ$C for approximately $10$ minutes.  The sample has been transferred to silicon for imaging. (c) AFM of the resulting structure.}\label{after}
\end{figure}

Figs.~\ref{before} and \ref{after} show the experiment in detail.  We use $3$~mm~$\times~12$~mm~$\times~70$~mm substrates of crosslinked polydimethylsiloxane (PDMS). Curing agent and PDMS prepolymer (Dow chemical Corning, Sylgard 184) are thoroughly mixed in a 1:30 weight ratio. After degassing, the polymer mixture is poured into a petri dish, cured at $85$~$^\circ$C for two hours and then left in the vacuum oven for the next 12-15 hours.  PS films are prepared by spin casting various concentrations of PS and toluene mixtures onto a freshly cleaved mica substrate.  The thickness of the film is controlled through the concentration of the PS solution.  We use PS of several molecular weights  (1.3 Mg/mol, 120 kg/mol Polymer Source, 106 kg/mol Fisher Scientific) and one polystyrene - b - poly(2-vinyl pyridine) molecule (symmetric diblock copolymer, 104 kg/mol, Polymer Source).  Finally, the prepared thin film is transferred to a clean deionized water surface (Milli-Q) and subsequently transferred to a PDMS substrate on a home built strain stage.  Samples are subsequently compressed and then relaxed as described above.  While the exact strain rate of the surface of the delamination was not measured, we estimate it to be $\sim 2 \times 10^{-3}$~s$^{-1}$ from the total time and the deformed geometry.  In all cases the films were imaged with Laser Scanning Confocal Microscopy (LSCM - Olympus Fluoview 1000) or with Atomic Force Microscopy (AFM - DI Dimension 2100) .

%\begin{table}
%\caption{Molecular Details: the average molecular weights and the polydispersity index for the polymer used in this study.}
%\label{MWTABLE}
%\begin{center}
%\begin{tabular}{lllll}
%\hline\noalign{\smallskip}
%Sample &   $f$ & $M_{W}$  (kg/mol) & $M_{N}$ (kg/mol) & PI \\
 %PS-1 & - &1260 & 1100 & 1.15 \\
 %PS-2 & - &119 & 115  & 1.04 \\
 %PS-3 & - &106 &  & 1.0 \\
 %PSPVP & 0.5 & 104 & 100 & 1.05 \\
%\end{tabular}
%\end{center}
%\end{table}

Fig~\ref{before}~b. shows a single LSCM optical section of a typical sample while under compression ($h=71$~nm).  The sample shows wrinkles (center), cracks (edge) and delamination (brightest wrinkle peak) as it has been compressed beyond the critical strain for wrinkling, $\epsilon _{w}$, cracking, $\epsilon_{c}$,  and delamination, $\epsilon_{D}$~\cite{Ebata2012}.  From such an image the wrinkle wavelength can easily be measured, but more importantly, through optical sectioning we also determine the height of each pixel.  Primarily, we use the reflection from the surface of the polymer (the highest intensity is recorded when the focal point of the lens is centered on the film surface allowing sub-Rayleigh resolution in the out of plane dimension).  Secondly, we verify the measurements by loading the film with a typical fluorescent molecule (Nile Red, Fisher Scientific).  Here the dye is excited with a $488$~nm excitation wavelength and imaged by a PMT behind a dichroic mirror which excludes the exciting laser light.  Finally, we also use AFM as a direct, high resolution measurement of the film topography (Fig~\ref{before}c.).  We therefore have a very careful and complete measurement of the film location in three dimensional space at all stages of compression and relaxation and can, for example, easily verify the predicted wrinkle, $f(x) = A \sin(kx)$, and delamination, $f(x) = (A/2)(1+\cos (2 \pi x/ w)$, topography ($f(x)$ is the surface, $A$ is an amplitude, $k$ is the wrinkle wavelength and $w$ is the width of the delamination)\cite{Bowden1998,Mei2011,Vella2009}.

\begin{figure}
\centering
\includegraphics[width=0.45\textwidth]{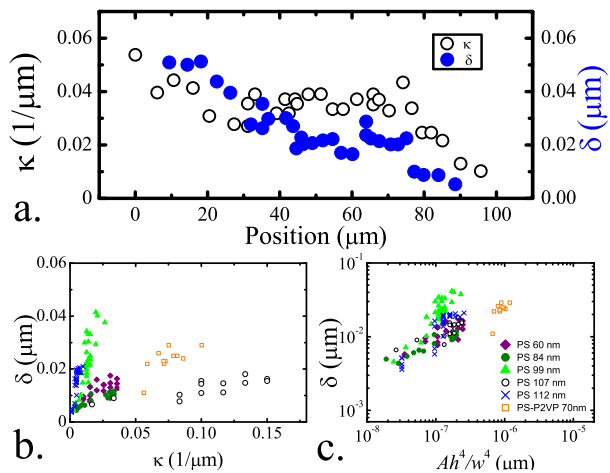}
\caption{(a) curvature and bump height as a function of position, illustrating the high degree of correlation between the two variables.  (b) Correlation plot of $\delta$ vs $\kappa$ showing several polymer films.  (c) Scaling plot showing collapse of the data.}. \label{bumps}
\end{figure}

After compression, a film is returned to its initially flat state (verified by LSCM and AFM, see Fig.~\ref{after}a.).  Aside from macroscopic fracture of the film, the sample shows no observable sign of damage.  The flat sample is then annealed above its \textit{bulk} glass transition temperature on a carefully controlled hot stage (Linkham, UK) and the surface temperature is measured to be $140 \pm 10$~$^\circ$C.  We observed a rapid development of thickness variation of height $\delta$ in the fluid polymer film (Fig.~\ref{after}b.).  The thickness variations 1) occur rapidly (see the supplementary file) 2) do not disappear on appreciable timescales (tens of minutes) and 3) correlate directly with regions of high amplitude during the compression stage of the experiment.  The timescales of emergence and decay of $\delta$ indicate that the driving force must greatly exceed the Laplace pressure that will eventually smooth the fluid surface, and because no hole opens to reveal the substrate $\delta$ is not related to dewetting~\cite{Reiter1994}.  The correlation between peak film curvature under compression and $\delta$ is evident upon comparison of the topography along the long axis of the feature, as shown in Fig.~\ref{bumps}a ($\kappa$ indicates curvature).  We explore the correlation for several different samples in Fig.~\ref{bumps}b., where we have recast curvature in terms of the surface strain, $\epsilon = \kappa h/2$ a more natural physical variable.  While each sample still shows a linear curve (correlation), we note that the samples do not collapse onto a master curve, and the slope does not show a monotonic dependence on sample thickness.  We show one curve generated with a thin diblock copolymer film (open squares) to highlight that the correlation is a general phenomena, and is not specific to PS.

In order to model the deformation upon heating we use the lubrication approximation to derive a more realistic scaling description of the film thickening.  We begin with a simplified Navier-Stokes equation, $\partial _t h(x,t) = (1/3 \eta) \partial_x (h^{3}_{0} \partial _{x} \sigma$), where $h_{0}$ is the initial film thickness, $\eta$ is the viscosity and $\partial$ represents a derivative with respect to the subscript variable.  $h(x,t)$ gives the film surface at a later time which, based on our observations, we assume to have the form  $h(x,t) = h_{0} + \delta(t) \cos (2 \pi x/w)$ where, $\delta(t)$, is the small perturbation we are interested in.  For simplicity, we will consider an independent delamination which has a curvature of $\kappa \sim (A \pi^2/w^2) \cos (2 \pi x/w)$.  The curvature leads to strain within the material of $\epsilon(z) = \kappa (z-h_{0}/2)$, where $z$ is normal to the substrate surface.    For simplicity, we assume a linear constitutive equation within the film $\sigma \sim  E_{f} \epsilon$, and that the stored strain is proportional to this bending stress.  Hence a perturbation of size $\delta$  to the film thickness scales as
\begin{equation}\label{delta}
\delta (t) \sim \frac{A h_{0}^{4} E_{f} }{\eta w^4},
\end{equation}
where we have assumed the time of observation, $t$, to be a constant as was the case in our experiments.  In Fig.~\ref{bumps}c. we show the scaled data.  The agreement with this simple model is remarkable given the small size of the thickened region ($\delta = 1 - 40$~nm) and the large exponents in the scaling rule.

While there is clearly some inaccuracy that needs to be overcome if Eqn.~\ref{delta} is to be used quantitatively (e.g. to measure viscosity or modulus of the thin film), its qualitative agreement gives confidence in the basic physics:  flow \textit{has} occurred in the fluidized film, and a stress \textit{must} have been stored in the film before it was annealed.  To make the measurement more quantitative we switch from considering the complete topography to considering only the critical point of the onset of plastic deformation (the yield strain).  This change significantly reduces measurement error because the focus is now a measurement of a large lengthscale (the distance from $\delta = 0$ to the sample edge, $\sim 50$~$\mu$m ).  A typical measurement as in the film shown in Fig.~\ref{before}, yields $\epsilon_p = 1.3\times 10^{-3}$ and reveals that the onset strain is an order of magnitude smaller than what might be expected from bulk measurements \cite{Argon1968, Kramer1974, Hasan1993}.  Our measurement, to the best of our knowledge, represents the first accurate measurement of the yield point in a thin polymer film~\footnote{While PS is well known to craze under tension, we were not able to observe any oriented structures with AFM. While this certainly does not mean that the film does not craze, we use the term yield due to a lack of evidence to the contrary.}.

\begin{figure}
\centering
\includegraphics[width=0.45\textwidth]{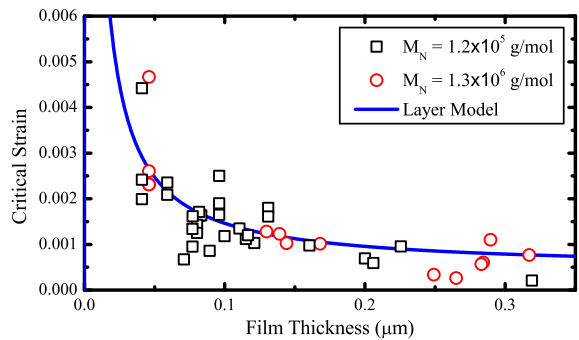}
\caption{Critical yield strain as a function of film thickness.  The figure shows data from two molecular weights differing by an order of magnitude and a fit to the layer model described in the text.   \label{confine}}
\end{figure}

We summarize the critical strain measurements by plotting $\epsilon_{p}$ as a function of film thickness in Fig.~\ref{confine}.  In addition to an overall small value of $\epsilon_{p}$, measurements show the critical strain \textit{increases} in films of thicknesses below $100$~nm.  While anomalous effects have often been measured in polymer films of this same thickness range \cite{Keddie1994,OConnell2005,Bodiguel2006,Reiter1994,Green2002,Ellison2003, Lee2012, Pye2011}, our measurement is novel in several respects.  First, this is a change in a heretofore untested material property relevant to failure, elastic theory and the basic sample preparation used in countless thin film experiments.  Secondly, this observation is a unique window on the behaviour of confined polymer systems from a mechanical perspective.  Thirdly, our experiments are independent of thermal strain (common to many thin film experiments) and because we observe no wrinkles after laminating the PS film to the PDMS substrate we can explicitly state a limit on any strain occuring due to the lamination of the thin film to the substrate (below $\sim 0.5 \%$).

The change we observe could be related to a reduction of inter-chain entanglements \cite{Brown1996, Si2005}, or to a change in the glass transition temperature of the film\cite{Keddie1994,Fakhraai2008, Ellison2003, Pye2011, OConnell2005, Bohme2002}.  Bending gives the opportunity to make a clear distinction between the two possibilities as it preferentially probes a film's free surface and the applied strains can be accurately measured in the vanishingly small strain limit.  Small strain ensures that the network of polymer chains is not deformed significantly, and therefore the entanglements themselves cannot play a role in the overall effect (this is not common in most failure measurements~\cite{Si2005, Lee2012}).  As evidence for this we duplicate the experiment with PS films of an order of magnitude larger molecular weight.  Fig.~\ref{confine} shows no observable effect of molecular weight on the location or magnitude of the upturn.  We tentatively conclude that the critical strain for the onset of plastic deformation must then be related to the same mechanism behind reductions in the glass transition.  While suggested by the simulations of B\"{o}hme and de Pablo, the relationship between the glass transition depression and plasticity has not yet been examined.  Qualitatively, the glass transition is reduced due to a region of increased mobility near the free surface of the polymer film (operating over a lengthscale of $\sim 10$~nm).  In this region a larger strain is needed before stress can be stored as stress can more easily relax away.  A trivial layered model can be written as $\epsilon_p (h) = (\ell / h) ( \epsilon _{p} - \epsilon _{p} ^{0}) + \epsilon _{p} ^{0}$, where $\ell$ is the size of the soft layer and $\epsilon _{p}$, $\epsilon _{p} ^{0}$ refer to the surface and bulk respectively.  A fit to our data in Fig.~\ref{confine}, assuming a typical lengthscale $\ell = 10$~nm, gives find $\epsilon _{p} = .017 \pm .002$ and $\epsilon _{p} ^{0} = .001 \pm .001$.

In conclusion, we have developed a sensitive new method to generate \textit{local} strain in thin polymer films and have used it to measure the onset of yield in polystyrene.  Our measurements show that the onset of yield occurs at much lower strains than are measurable in the bulk.  Furthermore, the critical strain shows significant increases in films thinner than $~100$~nm.  The confinement induced increases are independent of molecular weight, suggesting that they are related to changes in the glass transition temperature rather than changes in underlying inter-chain entanglement network.  Our measurement has implications for many current experimental investigations of the elastic properties of thin polymer films, particularly in experiments attempting to probe the far from threshold elastic behavior.

Funding for this work was provided by NSF-EPSCoR (EPS-0814442) and is gratefully acknowledged.  The authors would also like to thank Greg McKenna for pointing out some pertinent literature.

%\bibliography{TFB}

%pasted from .bbl file generated aug 28 2012.
%

\end{document}